\documentclass[12pt]{article}

\usepackage{graphicx}
\usepackage{amsmath}
\usepackage{amssymb}
\allowdisplaybreaks

\begin{document}

\baselineskip=17.5pt plus 0.2pt minus 0.1pt

\renewcommand{\theequation}{\arabic{equation}}
\renewcommand{\thefootnote}{\fnsymbol{footnote}}
\makeatletter
\def\CR{\nonumber \\}
\def\pt{\partial}
\def\eq#1{(\ref{#1})}
\def\la{\langle}
\def\ra{\rangle}
\def\hyp{\hbox{-}}


\begin{titlepage}
\title{Emergent general relativity \\ in 
fuzzy spaces from tensor models}
\author{
Naoki {\sc Sasakura}\thanks{\tt sasakura@yukawa.kyoto-u.ac.jp}
\\[15pt]
{\it Yukawa Institute for Theoretical Physics, Kyoto University,}\\
{\it Kyoto 606-8502, Japan}}
\date{}
\maketitle
\thispagestyle{empty}
\begin{abstract}
\normalsize
Tensor models can be regarded as theories of dynamical fuzzy spaces, and
provide background independent theories of space.
Their classical solutions correspond to classical background spaces,
and the small fluctuations around them can be regarded as fluctuations of fields on them.
In this paper, I numerically study a tensor model around its 
classical backgrounds of two-, three- and four-dimensional fuzzy flat tori and  
show that the properties of low-lying low-momentum modes are in clear agreement 
with general relativity. 
Numerical analysis also suggests that the lowest-order effective action 
is composed of curvature square terms, which is consistent with general relativity in 
view of the form of the considered action of the tensor model. 
\end{abstract}
\end{titlepage}

\section{Introduction}
\label{sec:intro}

Tensor models (or generalized matrix models) \cite{Ambjorn:1990ge}-\cite{Freidel:2005cg} 
were originally considered as the generalization of the matrix models,
which describe the two-dimensional simplicial quantum gravity, to higher-dimensional cases.
As in the matrix models, partition functions are given by a certain summation over
simplicial manifolds, which are dual to the Feynman diagrams. Although the matrix models  
have successfully been analyzed extensively by powerful analytical
methods, no analytical methods for tensor models are presently known.
Moreover, partition functions of tensor models lack physical interpretations
such as the topological sum over surfaces realized by the $1/N$-expansion in the matrix models.

In ref.~\cite{Sasakura:2005js}, 
a new interpretation of rank-three tensor models was proposed. This proposal is based 
on the facts that a fuzzy space is defined by an algebra of functions, and that 
an algebra can be characterized by a rank-three tensor that defines multiplication, 
$f_a f_b=C_{ab}{}^cf_c$. In this new interpretation that 
a rank-three tensor may be regarded as the structure constant of an algebra
defining a fuzzy space, rank-three
tensor models can be regarded as dynamical theories of fuzzy spaces.
This reinterpretation of the models is not just a formal difference, but provides a new
practical way of extracting physics from tensor models. Namely, this new interpretation makes
it meaningful to treat the models semiclassically, in which classical solutions of the models
can be regarded as background fuzzy spaces, and small fluctuations around solutions as
field fluctuations in fuzzy spaces. 
Another important difference from the original proposal \cite{Ambjorn:1990ge}-\cite{Freidel:2005cg}
is that three is enough as the rank of the tensor for describing a fuzzy space of any dimensions.
This property simplifies the structures of tensor models, since  
various dimensional cases can be considered in one form. 

The rank-three tensor models have mainly been analyzed by numerical means by the present author, 
and several interesting results have 
been obtained \cite{Sasakura:2005js, Sasakura:2005gv, Sasakura:2006pq, Sasakura:2007sv,
Sasakura:2007ud}. 
In particular, in the models that have a certain Gaussian type of classical solutions 
corresponding to fuzzy flat spaces of arbitrary dimensions,
it has numerically been shown that the properties of  
low-lying low-momentum small fluctuation modes around the Gaussian solutions are in good agreement
with the general relativity in the one-, two-, three- and four-dimensional cases
\cite{Sasakura:2007sv, Sasakura:2007ud}. In other words, general
relativity was found to emerge in these models as an effective low-momentum description. 
However, in the three- and four-dimensional cases, owing to the heavy computational requirement,
the agreement has been shown only for the numbers of the modes.
To conclude the agreement,
the main purpose of this paper is to carry out more detailed numerical analysis of 
the fluctuation modes 
in these dimensional cases by applying an approximate method similar
to that in ref.~\cite{Sasakura:2007sv} to the analysis of the tensor model introduced in
ref.~\cite{Sasakura:2007ud}, which is much simpler than that analyzed in ref.~\cite{Sasakura:2007sv}.

This paper is organized as follows. 
In the following section, the simpler model is reviewed. 
In \S\ref{sec:prediction}, by generalizing the discussions in ref.~\cite{Sasakura:2007ud}
to other dimensions,
the metric fluctuation modes in general relativity are studied. 
In \S\ref{sec:numerical}, the results of numerical analysis for the two-, 
three- and four-dimensional fuzzy tori are shown and compared with general relativity.
The final section is devoted to the conclusion and future problems. 

\section{The model}
\label{sec:model}
The tensor model that will be studied in this paper 
was introduced in my previous paper \cite{Sasakura:2007ud}. I summarize the main points below.

The dynamical variable is a totally symmetric rank-three real tensor:
\begin{align}
C_{abc}&:\mbox{real}, \\
\label{eq:csym}
C_{abc}&=C_{bca}=C_{cab}=C_{bac}=C_{acb}=C_{cba}.
\end{align} 
There is also a {\it nondynamical} symmetric rank-two real tensor:
\begin{align}
g^{ab}&:\mbox{real},\\
g^{ab}&=g^{ba},
\end{align}
which is basically taken to be $g^{ab}=\delta^{ab}$ (or $\delta(a-b)$ for continuous indices).
An action is assumed to have invariance under the orthogonal transformation $M_a{}^b$ as
\begin{equation}
\label{eq:cortho}
C_{abc}\rightarrow M_a{}^{a'}M_b{}^{b'}M_c{}^{c'} C_{a'b'c'}.
\end{equation}
This invariance can be achieved by contracting the upper and lower indices in pair, e.g.,
$g^{aa'}g^{bb'}g^{cc'}C_{abc}C_{a'b'c'}$. In some places, 
I will use $g^{ab}\ (g_{ab}\equiv (g^{-1})_{ab})$ 
to raise (lower) the indices of tensors.

The index of the tensors is the label of functions in a fuzzy space. 
It may take values in an infinite set or it may even become a continuous variable
in describing fuzzy spaces with an infinite number of functions such as a fuzzy infinite flat plane, 
or may take values in a finite set for a compact fuzzy space.  
The former case will appear below in defining a tensor model which possesses strictly Gaussian solutions.
The latter case will appear in \S\ref{sec:numerical}, 
where approximate Gaussian solutions corresponding to fuzzy flat tori are considered.
The former case may be considered as an infinite limit of the latter case.

At present, there are no principles for specifying an action. As an interesting case,
I consider an action that has the Gaussian-type solutions   
\begin{align}
\label{eq:cx}
\bar 
C_{x_1x_2x_3}&=B \exp\left[ -\beta \left( (x_1-x_2)^2+(x_2-x_3)^2+(x_3-x_1)^2\right)\right],\\
g^{x_1x_2}&=\delta^D(x_1-x_2), \nonumber
\end{align}
where $B$ and $\beta$ are positive numerical constants, 
$x_i$ are $D$-dimensional continuous coordinates,
$x_i=(x_i^1,x_i^2,\cdots,x_i^D)$, and  $(x)^2=\sum_{\mu=1}^D (x^\mu)^2$. 
The summation in the contraction of pairwise indices is defined as $\int d^Dx$. 
The algebra of functions $f_{x_1}f_{x_2}=\bar C_{x_1x_2}{}^{x_3}f_{x_3}$ defines
a commutative nonassociative fuzzy flat space considered in ref.~\cite{Sasai:2006ua}.
In the momentum basis, which is obtained by applying Fourier transforms to the indices, 
the solutions have the form
\begin{align}
\label{eq:cp}
\bar C_{p_1p_2p_3}&=A \,
\delta^D(p_1+p_2+p_3) \exp \left[ -\alpha \left( p_1^2+p_2^2+p_3^2\right) \right], \\
g^{p_1p_2}&=\delta^D(p_1+p_2),\nonumber
\end{align}
where $A$ and $\alpha= \frac{1}{12 \beta}$ are positive numerical constants. 
The summation in the contraction of pairwise indices is defined as
$\int d^Dp$.
The main reason for considering such solutions is the (partial) computability due to the Gaussian forms. 
Another reason is that the fuzzy spaces are physically well-behaved in the senses that their fuzziness
is well localized and that they have the Poincare symmetry. Moreover, as discussed 
in ref.~\cite{Sasakura:2007ud} and as will be reviewed in the next section, one can consider 
a natural correspondence between 
metric fluctuations in general relativity 
and fluctuations in tensor models around the Gaussian solutions.

There exist infinitely many actions that have the Gaussian solutions. An example was considered in 
ref.~\cite{Sasakura:2007sv}, and the counting of low-lying low-momentum modes showed agreement 
with general relativity in the one-, two-, three- and four-dimensional cases. 
A much simpler action was considered in ref.~\cite{Sasakura:2007ud}. 
This action contains a fractional power and cannot be considered to be well defined for 
arbitrary values of the tensors. However, at least for the small fluctuations around the 
Gaussian solutions, the action is well-defined and 
the properties of low-lying low-momentum modes showed clear agreement with general relativity 
in the two-dimensional case. In this paper, the analysis will be carried out for the much 
simpler action,
\begin{equation}
\label{eq:theaction}
S=W_{abc}W^{abc},
\end{equation}
where
\begin{align}
\label{eq:defwkh}
W_{abc}&=C_{abc}- 
(K^{-\frac29})_a{}^{a'}(K^{-\frac29})_{b}{}^{b'}(K^{-\frac29})_{c}{}^{c'}
H_{a'b'c'}, \\
\label{eq:defk}
K_a{}^{a'}&=C_{abc}C^{a'bc}, \\
H_{abc}&=C_{ade}C_{bf}{}^{d}C_c{}^{ef}.
\end{align}
When $A$ satisfies 
\begin{equation}
\label{eq:a}
A=\left(\frac{27 \alpha}{2\pi}\right)^\frac{D}{4},
\end{equation}
the Gaussian solutions \eq{eq:cp} satisfy
\begin{equation}
\label{eq:wzero}
W_{abc}=0,
\end{equation}
and become the classical solutions to the equation of motion from 
the action \eq{eq:theaction}.

\section{General relativity on flat tori}
\label{sec:prediction}
The strategy for comparing the tensor models with general relativity basically follows 
the procedure carried out for the two-dimensional case in my previous paper 
\cite{Sasakura:2007ud}.
Generalizing the Gaussian solutions \eq{eq:cx} in an invariant manner, 
the fluctuations of metric tensor in general relativity
and those of the tensor in tensor models are assumed to be related by\footnote{Here $g^{ab}$ of 
tensor models should not be confused with the metric tensor $g_{\mu\nu}(x)$.}
\begin{align}
\label{eq:correspondence}
C_{x_1x_2x_3}&=B g(x_1)^\frac14 g(x_2)^\frac14 g(x_3)^\frac14
\exp\left[ - \beta \left( d(x_1,x_2)^2+d(x_2,x_3)^2+d(x_3,x_1)^2\right)\right],\\
g^{x_1x_2}&=\delta^D(x_1-x_2), \nonumber
\end{align}
where $g(x)={\rm det}\left(g_{\mu\nu}(x)\right)$ and $d(x,y)$ denotes
the distance between two points, $x$ and $y$.

Although it is possible to directly compare by the correspondence \eq{eq:correspondence},
which should give the same final results, the actual comparison will be carried out in terms of 
the tensor $K_{ab}$ defined in \eq{eq:defk} for technical simplicity,
as in ref.~\cite{Sasakura:2007ud}. In tensor models, 
a small fluctuation, $\delta C_{abc}$, around a classical solution, $C^0_{abc}$, induces a fluctuation
of $K_{ab}$ expressed as
\begin{equation}
\label{eq:tensordk}
\delta K_{ab}=\delta C_{acd}\, {C^0{}_b}^{cd}+  C^0_{acd}\, {\delta C_b}^{cd}.
\end{equation}
On the other hand, the correspondence \eq{eq:correspondence} induces \cite{Sasakura:2007ud} 
\begin{equation}
\label{eq:metricdk}
\delta K_{p_1p_2}
=
\delta g_{\mu\nu}(p_1+p_2)\, 
(p_1-p_2)^\mu (p_1-p_2)^\nu \exp\left(-\frac1{16\beta}(p_1-p_2)^2 \right), 
\end{equation}
where the momentum basis is used, 
and $\delta g_{\mu\nu}(p)$ is the Fourier transform of $\delta g_{\mu\nu}(x)$,
which is the metric fluctuation around a flat background.
Equations \eq{eq:tensordk} and \eq{eq:metricdk} make it possible to compare the two
distinct types of fluctuation, $\delta C_{abc}$ in tensor models and $\delta g_{\mu\nu}(x)$
in general relativity.
 
In general relativity, not all of the components of $g_{\mu\nu}(x)$ 
are real geometric degrees of freedom on account of the gauge symmetry of local translations. 
As will be explained in the following section, the gauge symmetries in the tensor models 
appear as modes with vanishing spectra, 
and the modes to be compared are orthogonal to these gauge modes in 
the sense of the metric
\begin{equation}
\label{eq:Cmeasure}
ds_C^2=\delta C_{abc}\, \delta C^{abc},
\end{equation}
which is invariant under the orthogonal transformation \eq{eq:cortho}.
As reported in my previous paper \cite{Sasakura:2007ud}, 
by putting the correspondence \eq{eq:correspondence} into \eq{eq:Cmeasure}, 
one obtains
the DeWitt supermetric \cite{DeWitt:1962ud},
\begin{equation}
\label{eq:supermetric}
ds^2_{DW}=
 \int d^Dx \sqrt{g(x)} \left[
 \left(g^{\mu\nu}(x)\delta g_{\mu\nu}(x)\right)^2
+4 g^{\mu\rho}(x)g^{\nu\sigma}(x) \delta g_{\mu\nu}(x) 
\delta g_{\rho\sigma}(x)\right]. 
\end{equation}
Consequently, the metric fluctuation modes to be compared must 
be orthogonal to the local translational symmetry in the sense of 
the supermetric \eq{eq:supermetric}.

In the following numerical analysis, $D$-dimensional fuzzy flat tori will be
considered. Therefore, I will discuss below the orthogonal fluctuation modes in 
general relativity in the backgrounds of $D$-dimensional flat tori with metric 
$g_{\mu\nu}=\delta_{\mu\nu}$.
With such a background, the supermetric \eq{eq:supermetric} is explicitly given by
\begin{equation}
\label{eq:explicit}
ds^2_{DW}=
\int d^D x \Bigg[
\sum_{\mu=1}^D
5 (\delta g_{\mu\mu})^2+
\sum_{{\mu=1,\nu=2,}\atop{\mu<\nu}}^D \left\{ 2 \delta g_{\mu\mu}\delta g_{\nu\nu}+ 
8 (\delta g_{\mu\nu})^2\right\}
\Bigg].  
\end{equation}
The infinitesimal gauge transformation in the flat background is given by
\begin{equation}
\delta_v g_{\mu\nu}(x)=\partial_\mu v_\nu(x)+\partial_\nu v_\mu(x),
\end{equation}
where $v_\mu(x)$ is the local translation vector,
or 
\begin{equation}
\label{eq:gaugemom}
\delta_v g_{\mu\nu}(p)=i p_\mu v_\nu(p)+i p_\nu v_\mu(p)
\end{equation}
in the momentum basis.

As can be seen in \eq{eq:gaugemom},
at the vanishing momentum sector $p_\mu=0$, the gauge symmetry is null, and all the components
of the metric tensor are real geometric degrees of freedom.
Diagonalizing the supermetric \eq{eq:explicit}, the modes are classified into three orthogonal 
classes:
\begin{align}
\label{eq:zerodiag}
1\mbox{ conformal mode:}&\ \delta g_{\mu\mu}=\delta g_{\nu\nu}\neq 0 \mbox{ for all } \mu,\nu,
\mbox{ others}=0,\\
\label{eq:zerotraceless}
(D-1)\mbox{ traceless diagonal modes:}&
\ \sum_{\mu=1}^D \delta g_{\mu\mu}=0,\mbox{ off-diagonal components}=0, \\
\label{eq:zerooff}
D(D-1)/2 \mbox{ off-diagonal modes:}&\mbox{ diagonal components}=0.\ 
\end{align} 

At the nonvanishing momentum sector, one may take the momentum to be in the direction
$(p_1,0,\cdots,0)$ with evident generalization to the other directions.
Then the explicit form of the infinitesimal gauge transformation \eq{eq:gaugemom} is given by
\begin{align}
\label{eq:gaugenon}
\delta_v g_{11}&=2 i p_1v_1,\cr
\delta_v g_{1\mu}&=i p_1 v_\mu \mbox{ for }\mu\neq 1,\\
\delta_v g_{\mu\nu}&=0 \mbox{ for }\mu,\nu\neq 1.\nonumber
\end{align}
In the same manner,
the modes orthogonal to the gauge modes \eq{eq:gaugenon} 
in the sense of the supermetric \eq{eq:explicit} are
classified into three orthogonal classes:
\begin{align}
\label{eq:nonzeromixed}
1\mbox{ diagonal mode:}&\ \delta g_{11}=-D+1,\ \delta g_{\mu\mu}=5\mbox{ for all }\mu\neq 1, \\
\label{eq:nonzerotraceless}
(D-2)\mbox{ traceless diagonal modes:}&\ \delta g_{11}=0,\ \sum_{\mu=2}^D\delta g_{\mu\mu}=0, 
\mbox{ off-diag. comps.}=0, \\
\label{eq:nonzerooff}
(D-1)(D-2)/2\mbox{ off-diagonal modes:}&\ \delta g_{\mu\nu}\neq 0\mbox{ for }\mu\neq \nu,\ \mu,\nu
\neq 1.
\end{align}

\section{Numerical results}
\label{sec:numerical}

I will first review the general strategy for obtaining the quadratic potential 
for small fluctuations in tensor models, 
which was carried out in refs.~\cite{Sasakura:2007sv, Sasakura:2007ud}.

The components of $C_{abc}$ in the tensor models are subject to the symmetry 
constraint \eq{eq:csym}.
The independent normalized fluctuation components are obtained from the metric 
\eq{eq:Cmeasure} as
\begin{equation}
\label{eq:Cmeasuresym}
ds^2_C=\sum_{(a,b,c)} m[(a,b,c)]\, 
\delta C_{abc}\delta C^{abc}=\sum_{(a,b,c)} \delta \tilde C_{(a,b,c)}
\delta \tilde C^{(a,b,c)},
\end{equation}
where $(a,b,c)$ denotes an order-independent set of $a,b$ and $c$; $\tilde C_{(a,b,c)}$ are 
the normalized independent components defined by
$\tilde C_{(a,b,c)}=\sqrt{m[(a,b,c)]}\,C_{abc}$; 
and $m[(a,b,c)]$ is the multiplicity defined by
\begin{equation}
\label{eq:multifac}
m[(a,b,c)]=\left\{
\begin{array}{cl}
1 & a=b=c, \\
3 & a=b\neq c,\ b=c\neq a, {\rm \ or\ } c=a \neq b,\\
6 & {\rm otherwise}.
\end{array}
\right.
\end{equation}
Then the coefficient matrix of the quadratic potential for the normalized small fluctuation
components
around the Gaussian solutions \eq{eq:cp} with \eq{eq:a} is given by
\begin{equation}
\label{eq:M}
M^{(a,b,c),(d,e,f)}=\sum_{g,h,i}
\frac{\partial W_{ghi}}{\partial \tilde C_{(a,b,c)}} 
\frac{\partial W^{ghi}}{\partial \tilde C_{(d,e,f)}}\Bigg|_{C=\bar C}
=\sum_{g,h,i}\frac{1}{\sqrt{m[(a,b,c)]m[(d,e,f)]}}
\frac{\partial W_{ghi}}{\partial C_{abc}} 
\frac{\partial W^{ghi}}{\partial C_{def}}\Bigg|_{C=\bar C}, 
\end{equation}
where I have used the form of the action \eq{eq:theaction} and the fact that \eq{eq:wzero} is 
satisfied by the Gaussian solutions \eq{eq:cp} with \eq{eq:a}. 

The numerical analysis will be carried out around the backgrounds of $D$-dimensional fuzzy flat tori.
These are the simplest fuzzy spaces with a finite number of functions on it and are suited for
numerical and analytical analyses. Because of the translational invariance of the backgrounds, 
it is convenient to take momentum bases, where an index is assumed to take discrete values in a finite
range as
\begin{equation}
\label{eq:prange}
p=(p^1,p^2,\ldots,p^D),\ \ (p^i=-L,-L+1,\cdots,L).
\end{equation}

The Gaussian solutions \eq{eq:cp} with evident replacement of the $\delta$-function 
with a Kronecker delta will be the solutions to the action 
\eq{eq:theaction} in the limit $L\rightarrow \infty$ with $1/L^2 \lesssim \alpha \lesssim 1/L$,
since the effects caused by the finiteness and discreteness of momenta will 
vanish in this limit\footnote{In this limit, the physical size of a torus 
becomes infinitely large in comparison with the scale of fuzziness $\sqrt{\alpha}$.} 
\cite{Sasakura:2007sv}. In fact, in my previous paper \cite{Sasakura:2007ud},
correct numerical solutions similar to the Gaussian solutions \eq{eq:cp} 
were explicitly found for the action \eq{eq:theaction} with such discrete and finite momentum indices 
in the two-dimensional case. 
Therefore, it can be expected that the coefficient matrix \eq{eq:M} with \eq{eq:prange} will 
give a good approximation for $D$-dimensional fuzzy flat tori, 
when $L$ is large enough and $\alpha$ is in the above range.

The explicit form of the partial derivative in \eq{eq:M} was given in \cite{Sasakura:2007ud} in
the momentum basis as   
\begin{equation}
\frac{\partial W_{abc}}{\partial C_{def}}\Bigg|_{C=\bar C}=\frac{m[(d,e,f)]}{(3!)^2}\sum_{\sigma,\sigma'}
T_{\sigma(a)\sigma(b)\sigma(c)}^{\sigma'(d)\sigma'(e)\sigma'(f)},
\end{equation}
where the summations are over all the permutations of the indices, and
\begin{align}
T_{p_1p_2p_3}^{p_4p_5p_6}=&
\delta_{p_1}^{p_4} \delta_{p_2}^{p_5}\delta_{p_3}^{p_6}
-3 \delta_{p_1+p_2}^{p_4+p_5}\delta_{p_3}^{p_6}  (k_{p_1}k_{p_2}k_{p_3})^{-\frac29} c_{p_1,-p_4,-p_1+p_4}
c_{-p_5,p_2,-p_2+p_5} \cr
&+3 \delta_{p_1}^{p_4+p_5} \delta_{p_2+p_3}^{p_6} (k_{p_1})^{-\frac29} f_{p_1,p_2+p_3}c_{p_1,-p_4,-p_5}
c_{p_2,p_3,-p_6}\cr
&+
3 \delta_{p_1}^{p_4} \delta_{p_2+p_3}^{p_5+p_6} (k_{p_1})^{-\frac29} f_{p_1,p_2+p_3} c_{p_2,p_3,-p_2-p_3}
c_{p_5+p_6,-p_5,-p_6},\\
c_{p_1,p_2,p_3}=& A \exp [-\alpha(p_1^2+p_2^2+p_3^2)], \\
k_p=& A^2 \left(\frac{\pi}{4\alpha}\right)^{\frac{D}2} 
\exp\left(-3 \alpha p^2 \right),\\
f_{p_1,p_2}=&\left\{
\displaystyle 
\begin{array}{l}
\frac{(k_{p_1})^\frac29-(k_{p_2})^\frac29}{k_{p_1}-k_{p_2}}\mbox{ for }k_{p_1}\neq k_{p_2},\\
\frac{2}{9} (k_{p_1})^{-\frac79} \mbox{ for } k_{p_1} = k_{p_2}
\end{array}
\right.
,
\end{align}
with $A$ given by \eq{eq:a}.

In general, the coefficient matrix \eq{eq:M} contains a number of vanishing eigenvalues for 
a nontrivial classical background solution. These zero modes come from the broken generators of
the orthogonal symmetry \eq{eq:cortho} in such a background, and have been 
analyzed completely for $D$-dimensional fuzzy flat tori in ref.~\cite{Sasakura:2007sv}.
Since these modes should physically correspond to local gauge symmetry
nonlinearly realized in a symmetry-breaking background \cite{Borisov:1974bn}, 
they are not of interest in comparison with the orthogonal metric fluctuation modes 
discussed in the previous section.
In the present approximate treatment, 
these modes generally have very small but nonvanishing eigenvalues
and cannot be distinguished from the modes of interest particularly in the low-momentum regions,
as was problematic in ref.~\cite{Sasakura:2007sv}.
To overcome the problem, in the present numerical analysis, the coefficient matrix \eq{eq:M} 
will be projected to the subspace
orthogonal\footnote{In the sense of the metric \eq{eq:Cmeasure}.} to the approximate 
broken symmetry modes that are obtained by infinitesimally applying the  orthogonal transformations 
\eq{eq:cortho} to 
the Gaussian solutions \eq{eq:cp}. The eigenvalue/mode analysis will be carried out
for the projected coefficient matrix. 

Because of the translational symmetry of the torus backgrounds,
the coefficient matrix \eq{eq:M} is a direct sum of the submatrices of each momentum sector
labeled by the momentum $p=p_1+p_2+p_3$ of the 
fluctuations $\delta C_{p_1p_2p_3}$. Therefore
the eigenvalue/mode analysis will be carried out for each momentum sector
of the fluctuations.

The numerical analysis is performed on a Windows XP64 workstation
with two Opteron 275 (2.2 GHz, dual core each) processors and 8 GB memories. 
The C++ codes\footnote{Interested readers can download the C++ codes from 
http://www2.yukawa.kyoto-u.ac.jp/\~\ sasakura/codes/gausscodes.zip .} are
compiled by the Intel C++ compiler 10.1 with OpenMP parallelization. 
ACML 4.0 is used for LAPACK/BLAS routines. 
Mathematica 6.0 is used for analyzing the outputs.

\begin{figure} 
\begin{center}
\includegraphics[scale=.9]{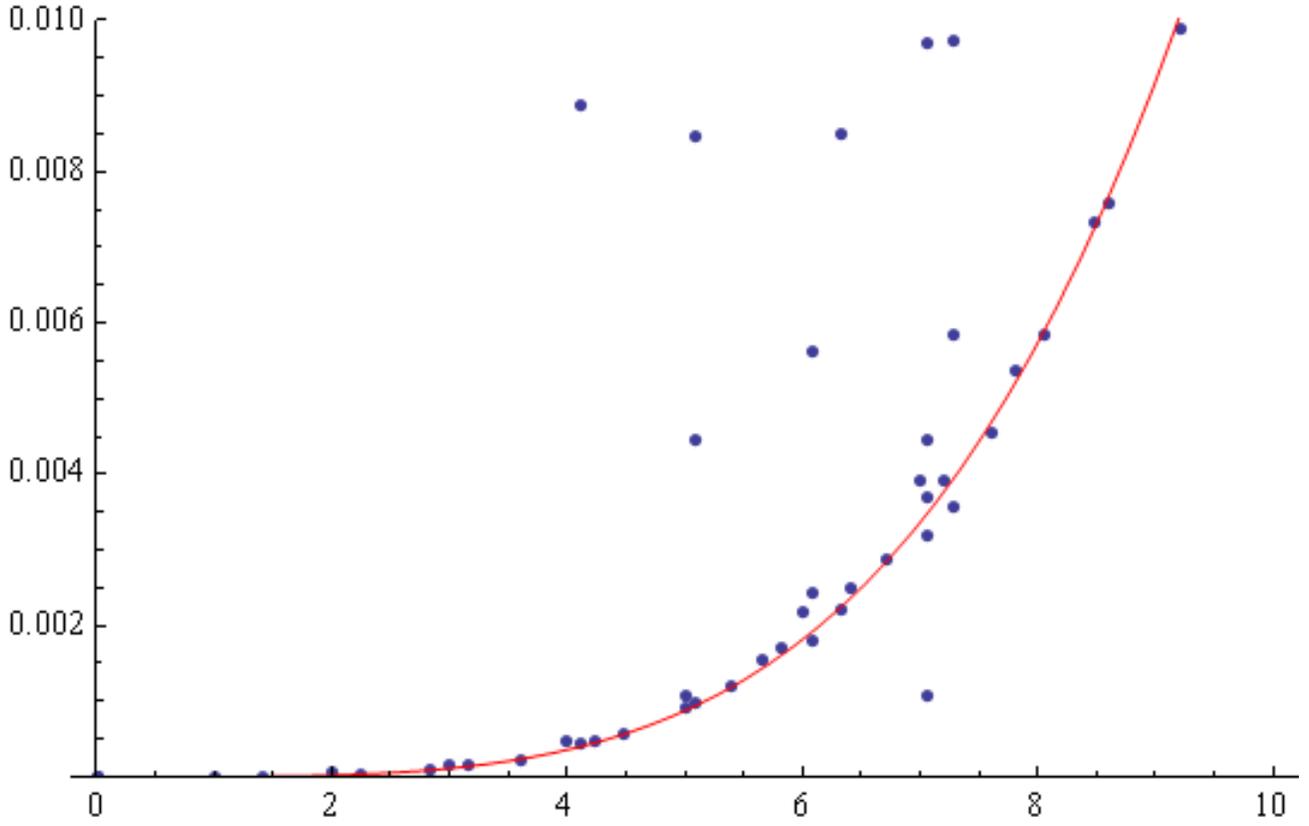}
\end{center}
\caption{Low-lying spectra on two-dimensional fuzzy torus for $L=10$ and $\alpha=1.5/L^2$. The horizontal 
axis is the size of momentum $|p|=\sqrt{(p^1)^2+(p^2)^2}$ of each momentum sector, and the vertical axis
is the spectral value. The solid line is $1.4\times 10^{-6} |p|^4$.}
\label{fig:dim2L10spec}
\end{figure}
\begin{figure} 
\begin{center}
\includegraphics[width=5cm]{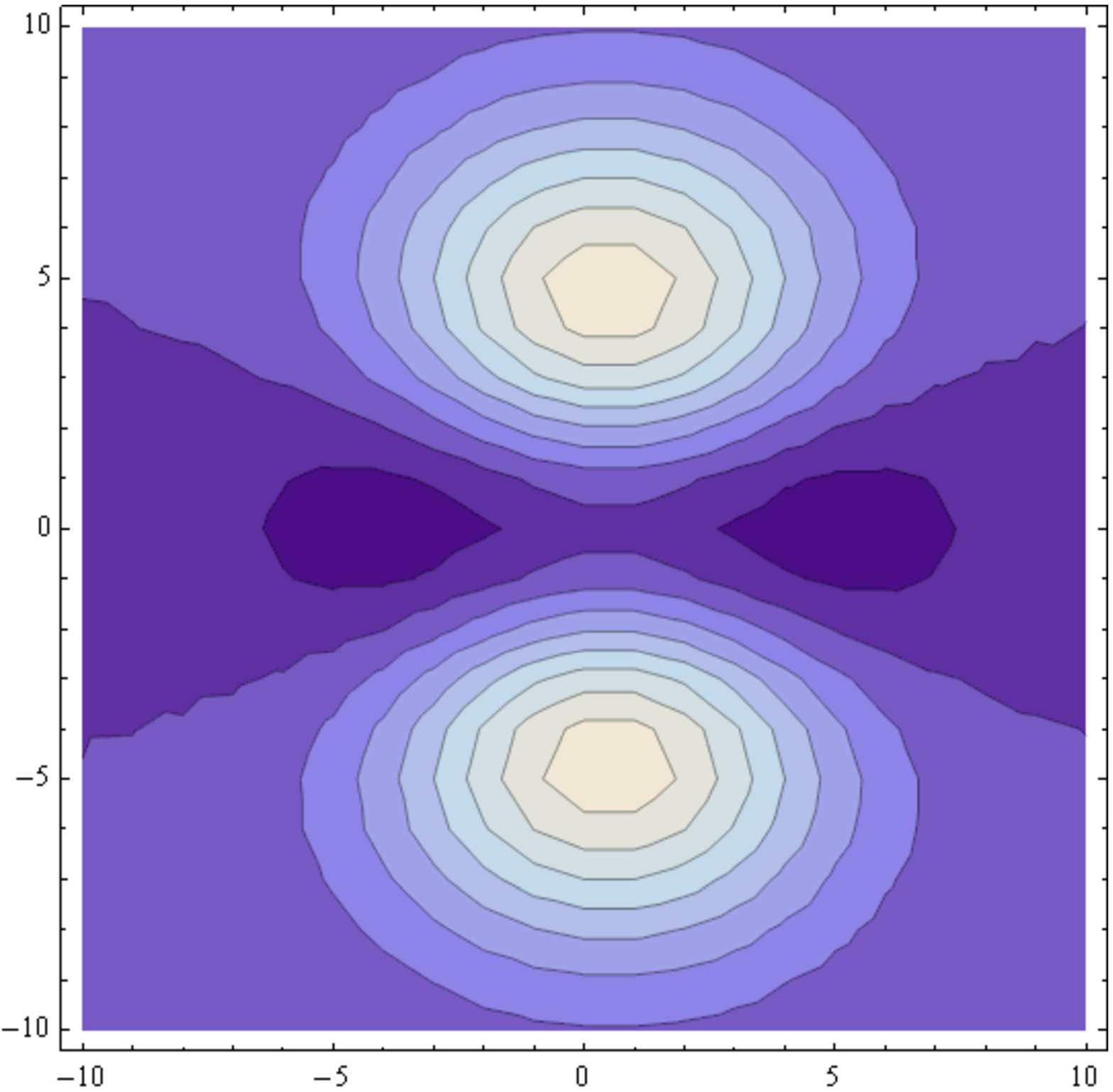}
\hspace{2cm}
\includegraphics[width=5cm]{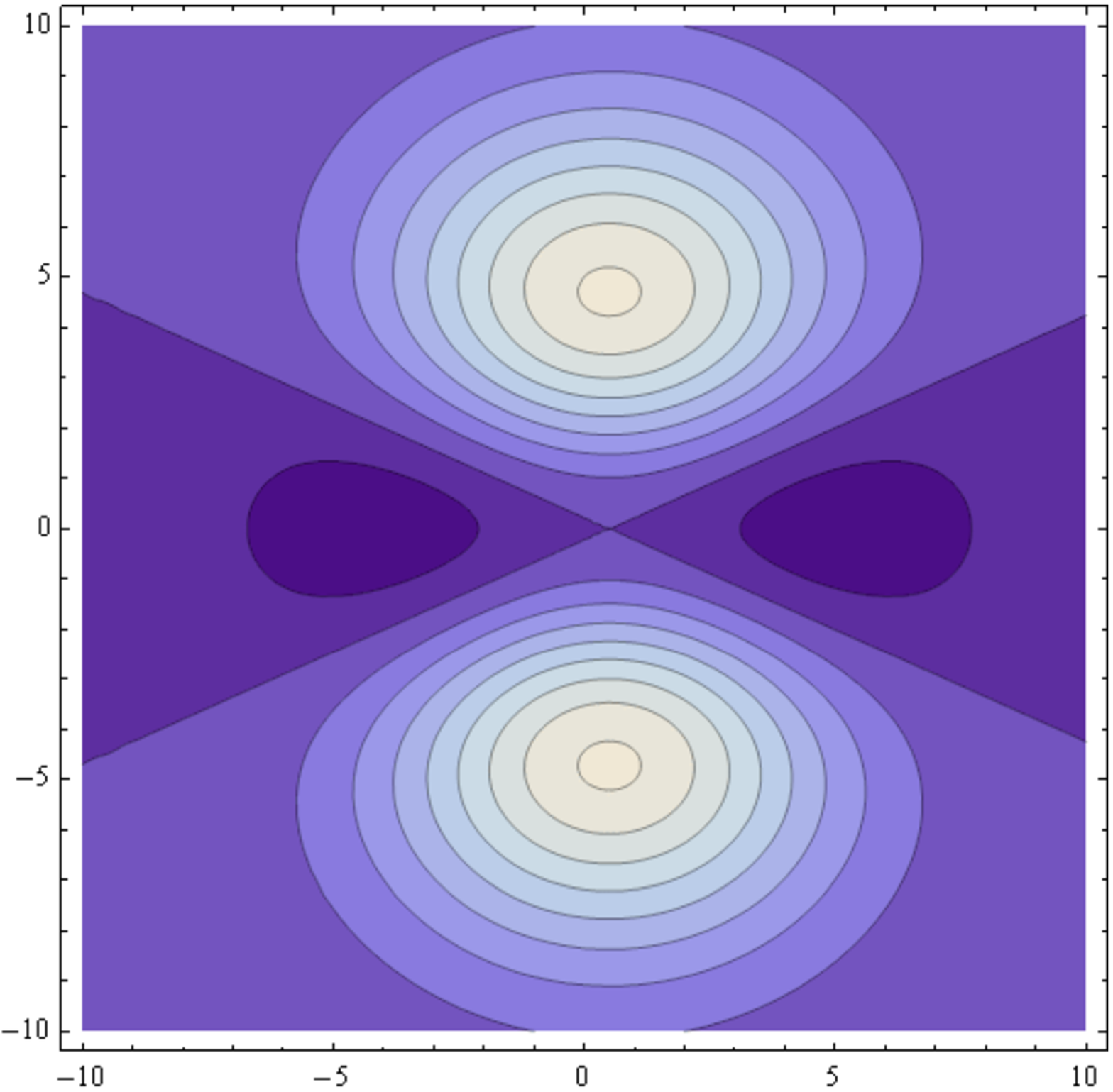}
\end{center}
\caption{Contour plots of $\delta K_{q,-q+p}$ for low-lying mode in $p=(1,0)$ sector
in two-dimensional case.
The left- and right-hand figures are \eq{eq:tensordk} from the numerical computations
and \eq{eq:metricdk} substituted with \eq{eq:nonzeromixed}, respectively.
The axes are $q=(q^1,q^2)$.}
\label{fig:dim2L10}
\end{figure}
Let me first show the results for the two-dimensional case. In my previous paper \cite{Sasakura:2007ud},
an eigenvalue/mode analysis was carried out 
in a numerically correct way without using the 
approximation explained above. The merit of the approximation is that a larger $L$
can be taken owing to the efficiency of computation.
In Fig.~\ref{fig:dim2L10spec}, the case of $L=10$ and $\alpha=1.5/L^2$ is shown. 
In the figure, a trajectory of spectra can be clearly 
seen up to $|p|\lesssim 6$, 
and no overlap with some unwanted spectra at $p=(2,1)$ for $L=5$ reported in my previous paper 
\cite{Sasakura:2007ud} is observed. This supports the expectation that, 
as $L$ increases, the agreement with general relativity becomes clearer. 
The spectra are well located on the massless trajectory of $|p|^4$, 
which is in agreement with the result of \cite{Sasakura:2007ud}.   

As reported in ref.~\cite{Sasakura:2007ud}, 
the properties of the low-lying modes are in clear agreement with 
the results from general relativity in the previous section. 
In the $p=0$ sector, there are three low-lying modes with the spectra 
$1.6\times 10^{-6}$, $6.4\times 10^{-6}$ and $6.6\times 10^{-6}$. 
By comparing the numerical results of \eq{eq:tensordk}
and the analytical expression \eq{eq:metricdk}, these modes were clearly identified with 
\eq{eq:zerooff}, \eq{eq:zerodiag} 
and \eq{eq:zerotraceless}, respectively.  
The low-lying mode at each nonvanishing momentum sector 
was clearly consistent with \eq{eq:nonzeromixed}, 
as is shown for $p=(1,0)$ in Fig.~\ref{fig:dim2L10}.

\begin{figure} 
\begin{center}
\includegraphics[scale=.9]{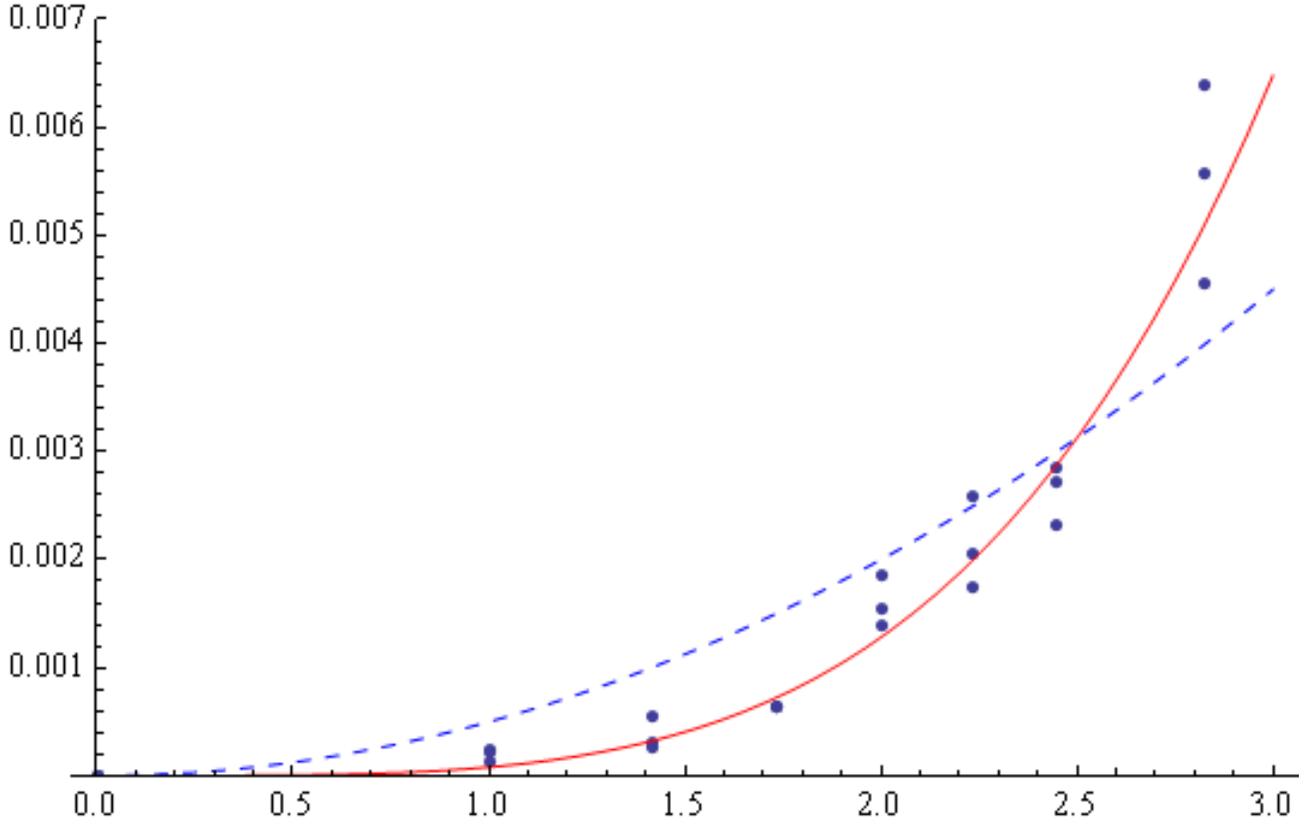}
\end{center}
\caption{Low-lying spectra on three-dimensional fuzzy torus for $L=3$ and $\alpha=1/L^2$. 
The solid and dashed lines are $8\times 10^{-5} |p|^4$ and $5\times 10^{-4} |p|^2$, respectively.}
\label{fig:dim3L3spec}
\end{figure}
\begin{figure} 
\begin{center}
\includegraphics[width=3.9cm]{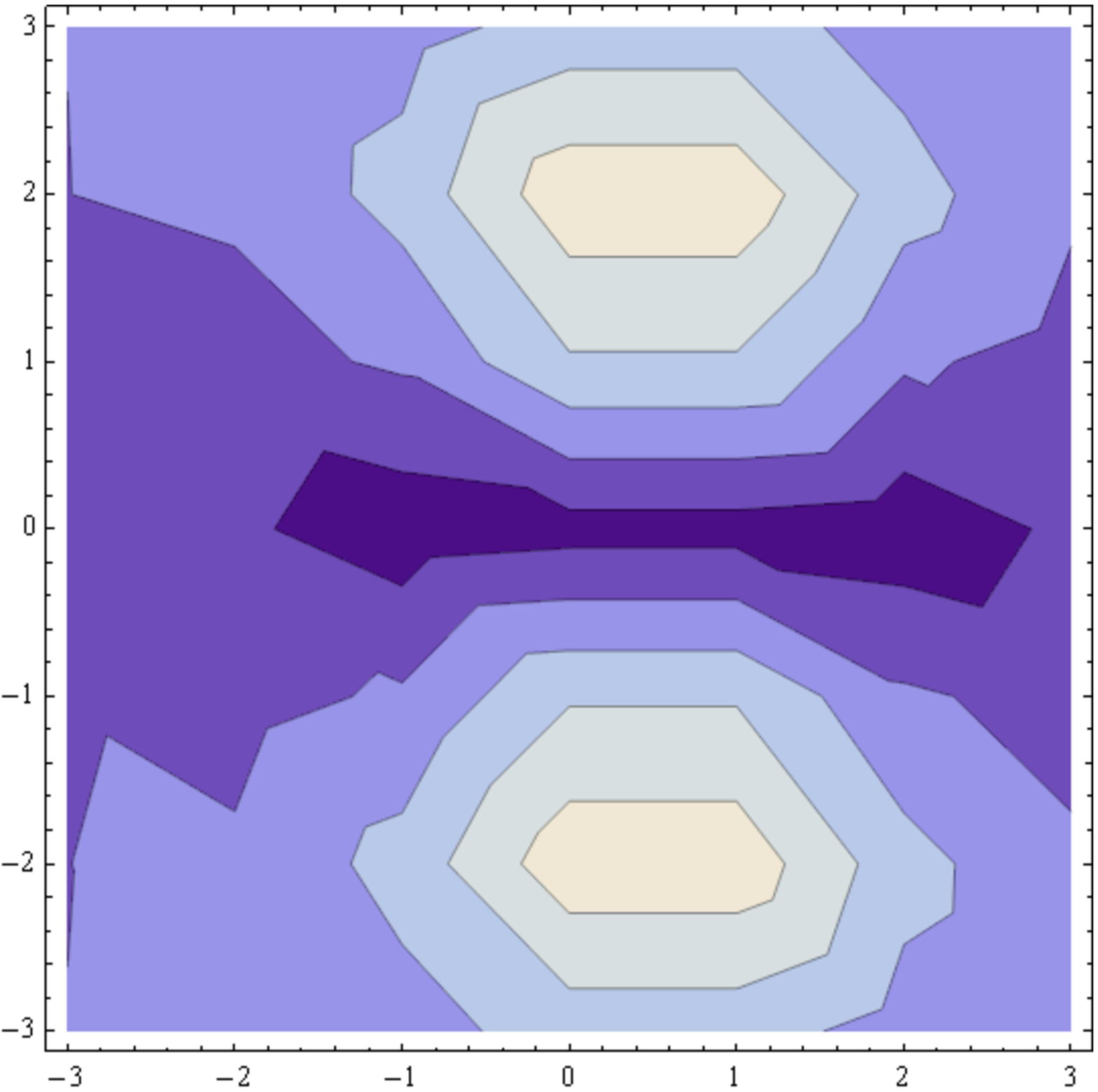}
\includegraphics[width=3.9cm]{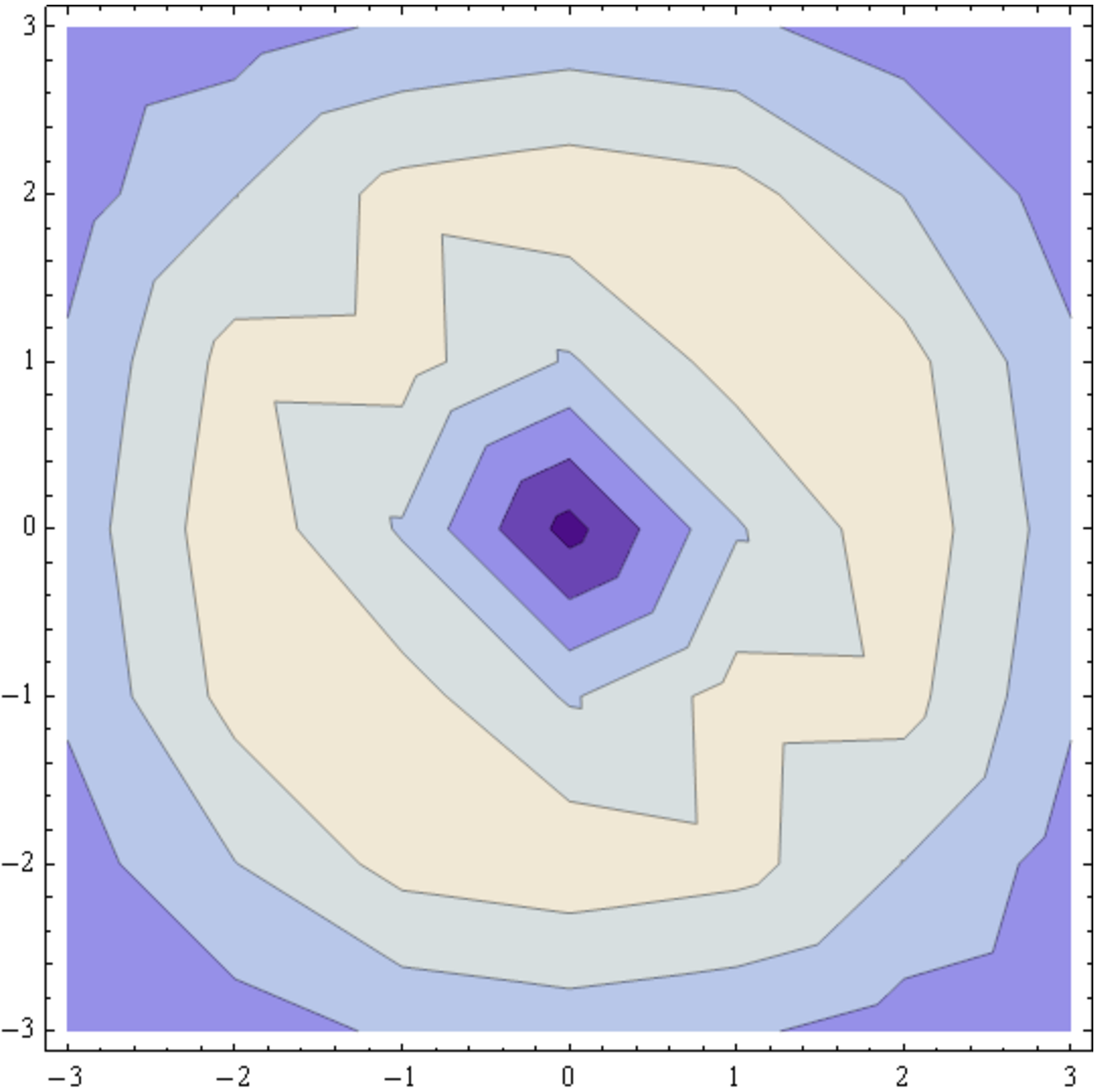}
\hspace{.5cm} 
\includegraphics[width=3.9cm]{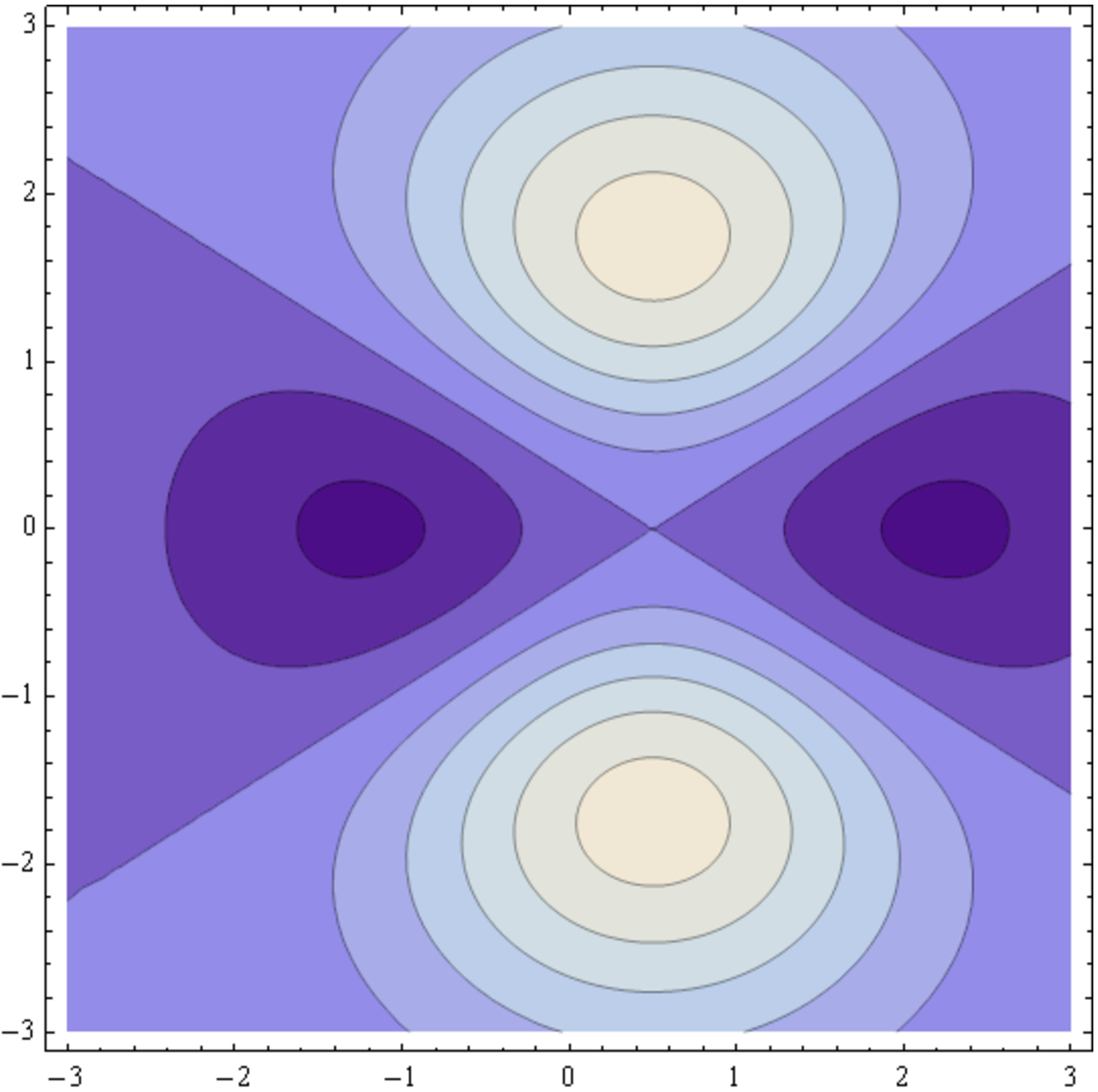}
\includegraphics[width=3.9cm]{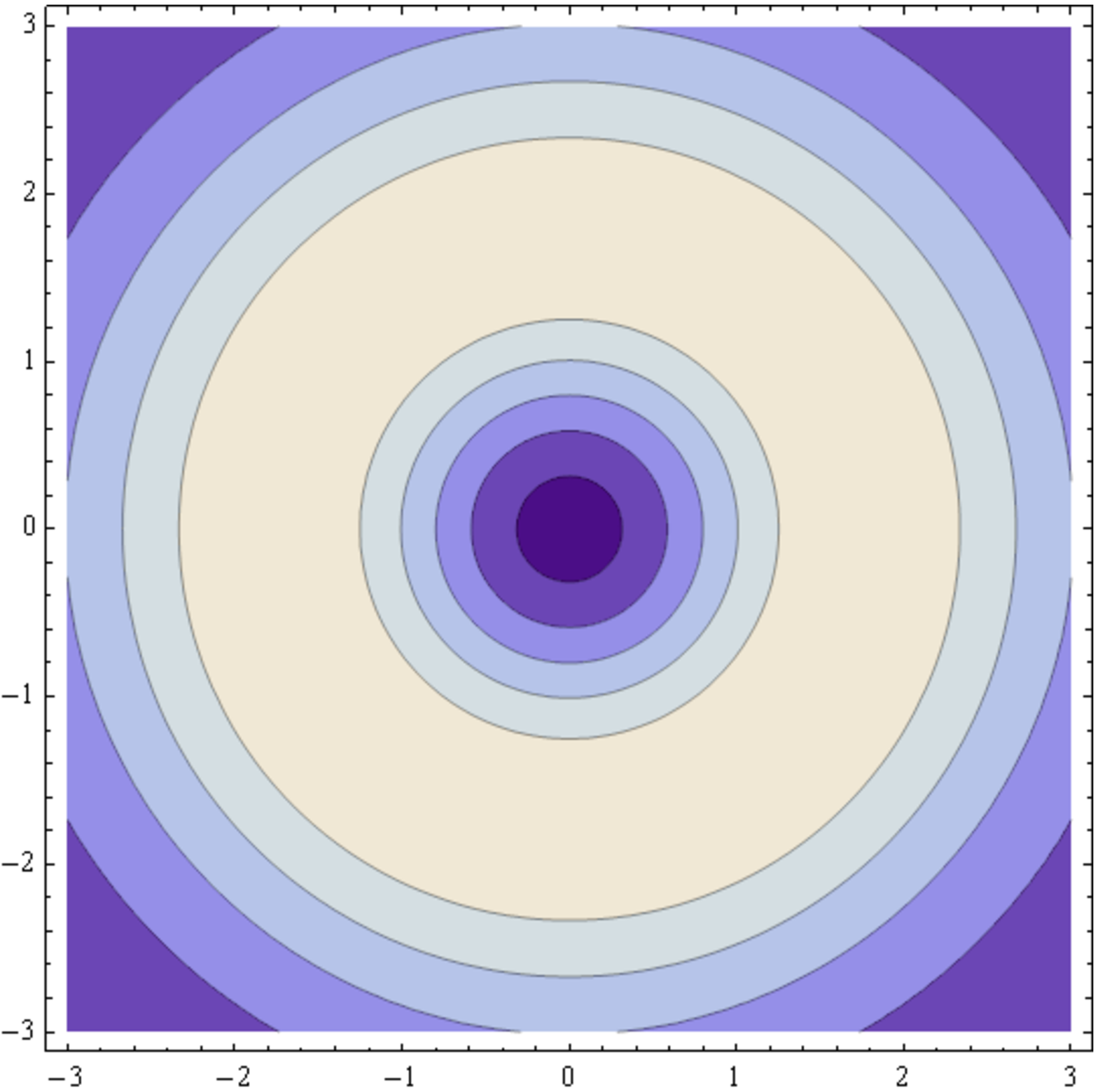}
\end{center}
\caption{Contour plots of $\delta K_{q,-q+p}$ for second low-lying mode in $p=(1,0,0)$ 
sector in three-dimensional case.
The left- and right-hand pairs of figures show \eq{eq:tensordk} from the numerical computations 
and \eq{eq:metricdk} substituted with \eq{eq:nonzeromixed}, respectively.
The left- and right-hand figures in each pair show 
$\delta K_{q,-q+p}$ at the slices $q=(q^1,q^2,0)$ and 
$q=(0,q^2,q^3)$, respectively.}
\label{fig:dim3L3}
\end{figure}
In the three-dimensional case, I will show the numerical results for $L=3$ and $\alpha=1/L^2$.
This number of $L$ is the maximum value feasible within the machine memory size.
In Fig.~\ref{fig:dim3L3spec}, the low-lying spectra are plotted.  
$|p|^4$ seems to be better than $|p|^2$ as a fitting line for the spectra. 

In more detail, in the $p=0$ sector, there are three types of spectrum, $3.4\times 10^{-6}$,
$5.4\times 10^{-6}$ and $7.7\times 10^{-6}$ with degeneracies of two, three, and one,
respectively. Therefore, these modes can be consistently identified with the modes 
\eq{eq:zerotraceless}, \eq{eq:zerooff} and \eq{eq:zerodiag}, respectively. 
In fact, the fitting of \eq{eq:metricdk} to \eq{eq:tensordk} of each mode from the numerical computations
was clearly consistent with the identification. 
For example, fitting to the last mode resulted in $\delta g_{\mu\mu}=2.45\pm 0.02,\ 
\delta g_{\mu\nu}=0\pm 0.01$, which is in clear agreement with \eq{eq:zerodiag}.

In each $p\neq 0$ sector, there are three low-lying modes. This number agrees with the 
result in the previous section. For example, 
the $p=(1,0,0)$ sector has three low-lying spectra, $1.4\times 10^{-4}$, $2.1\times 10^{-4}$ and 
$2.3\times 10^{-4}$. By a similar fitting, they can be clearly identified with the modes 
\eq{eq:nonzerooff}, \eq{eq:nonzeromixed} and \eq{eq:nonzerotraceless}, respectively.
For example, an analysis of the second mode resulted in
$\delta g_{11}=-0.74 \pm 0.01,\ \delta g_{22,33}=1.88 \pm .01,\ \delta g_{\mu\nu}=0\pm 0.01$, 
which is in clear agreement with \eq{eq:nonzeromixed}.
In Fig.~\ref{fig:dim3L3}, \eq{eq:tensordk} for the second mode
and \eq{eq:metricdk} substituted with \eq{eq:nonzeromixed} are shown. 

\begin{figure} 
\begin{center}
\includegraphics[scale=.9]{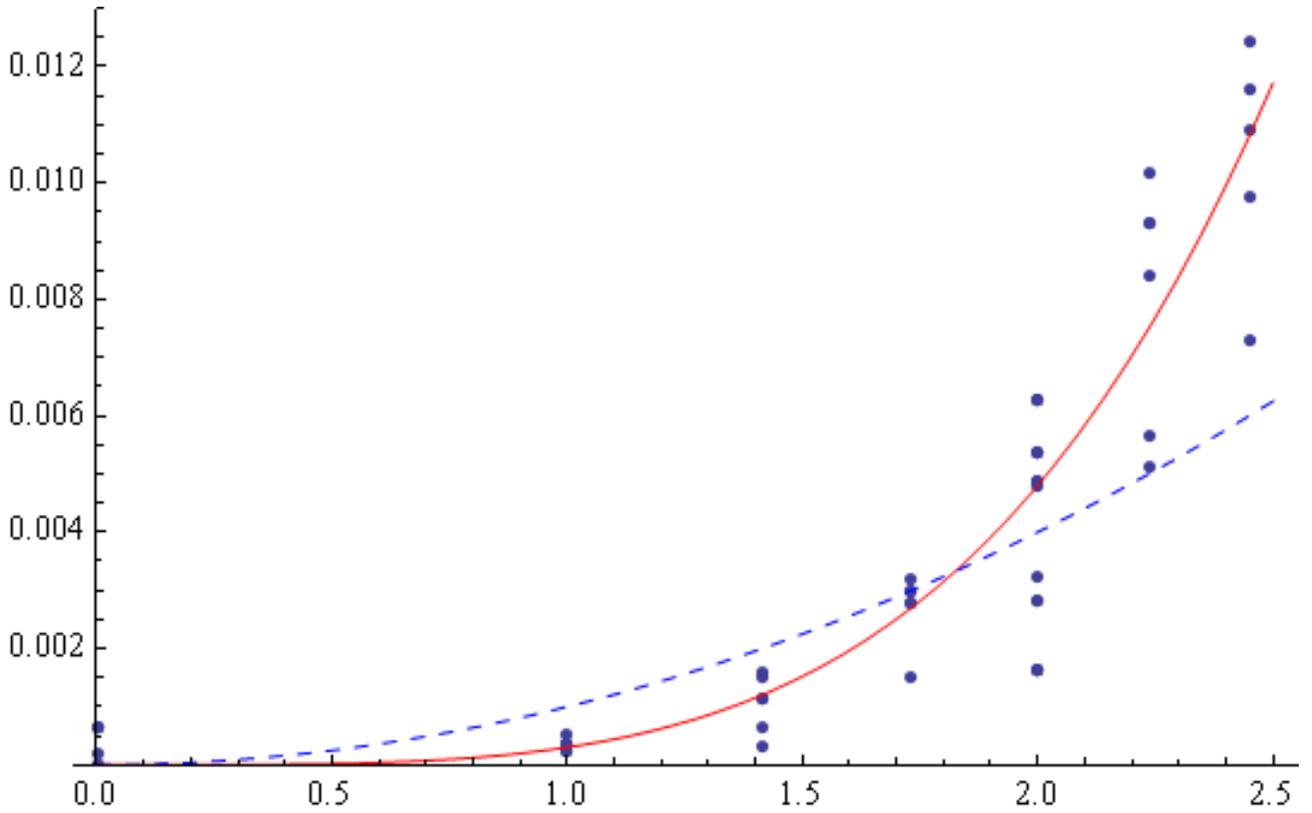}
\end{center}
\caption{Low-lying spectra on four-dimensional fuzzy torus for $L=2$ and $\alpha=1/L^2$. 
The solid and dashed lines are $3\times 10^{-4} |p|^4$ and $1\times 10^{-3} |p|^2$, respectively.}
\label{fig:dim4L2spec}
\end{figure}
\begin{figure} 
\begin{center}
\includegraphics[width=3.9cm]{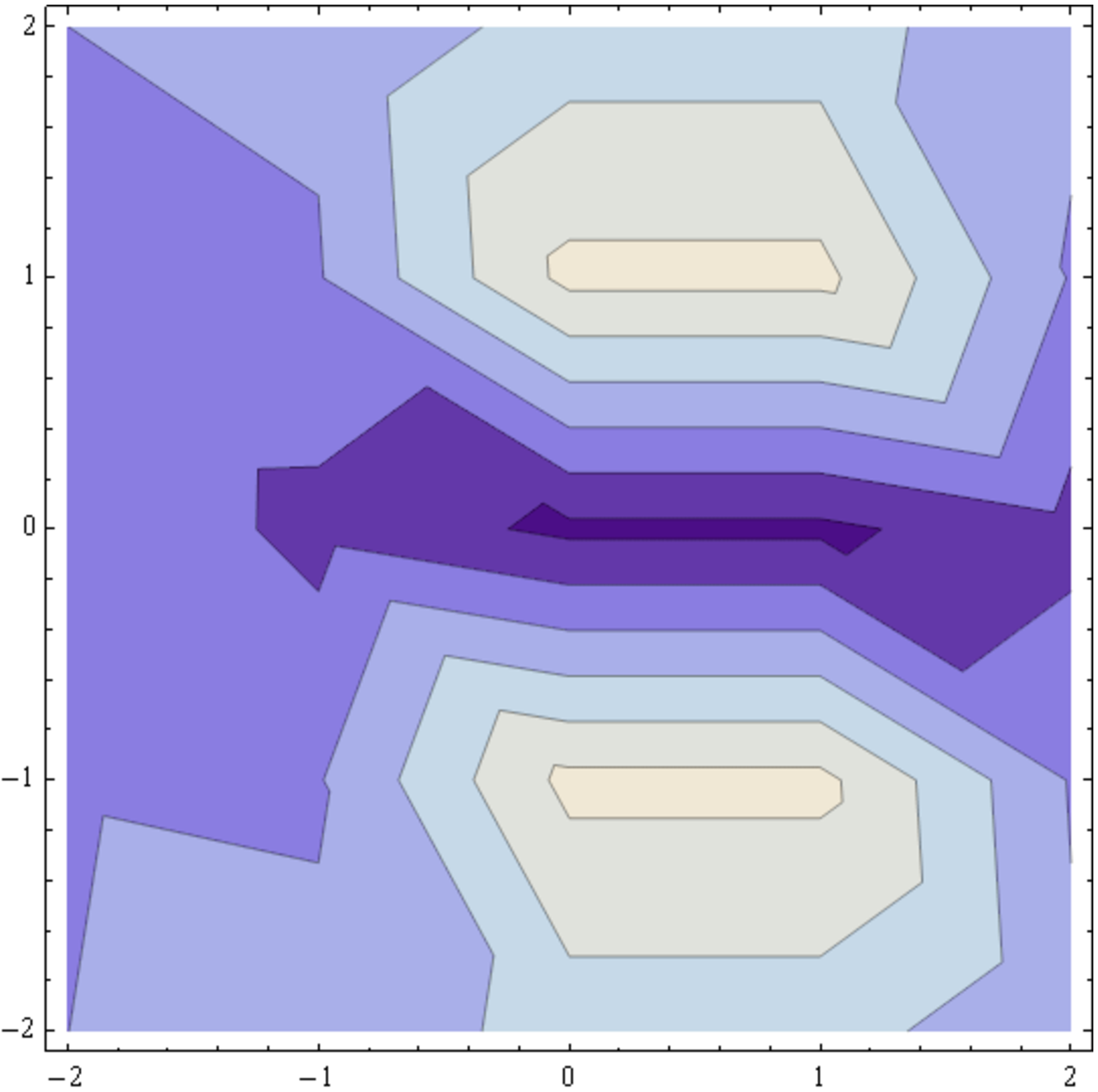}
\includegraphics[width=3.9cm]{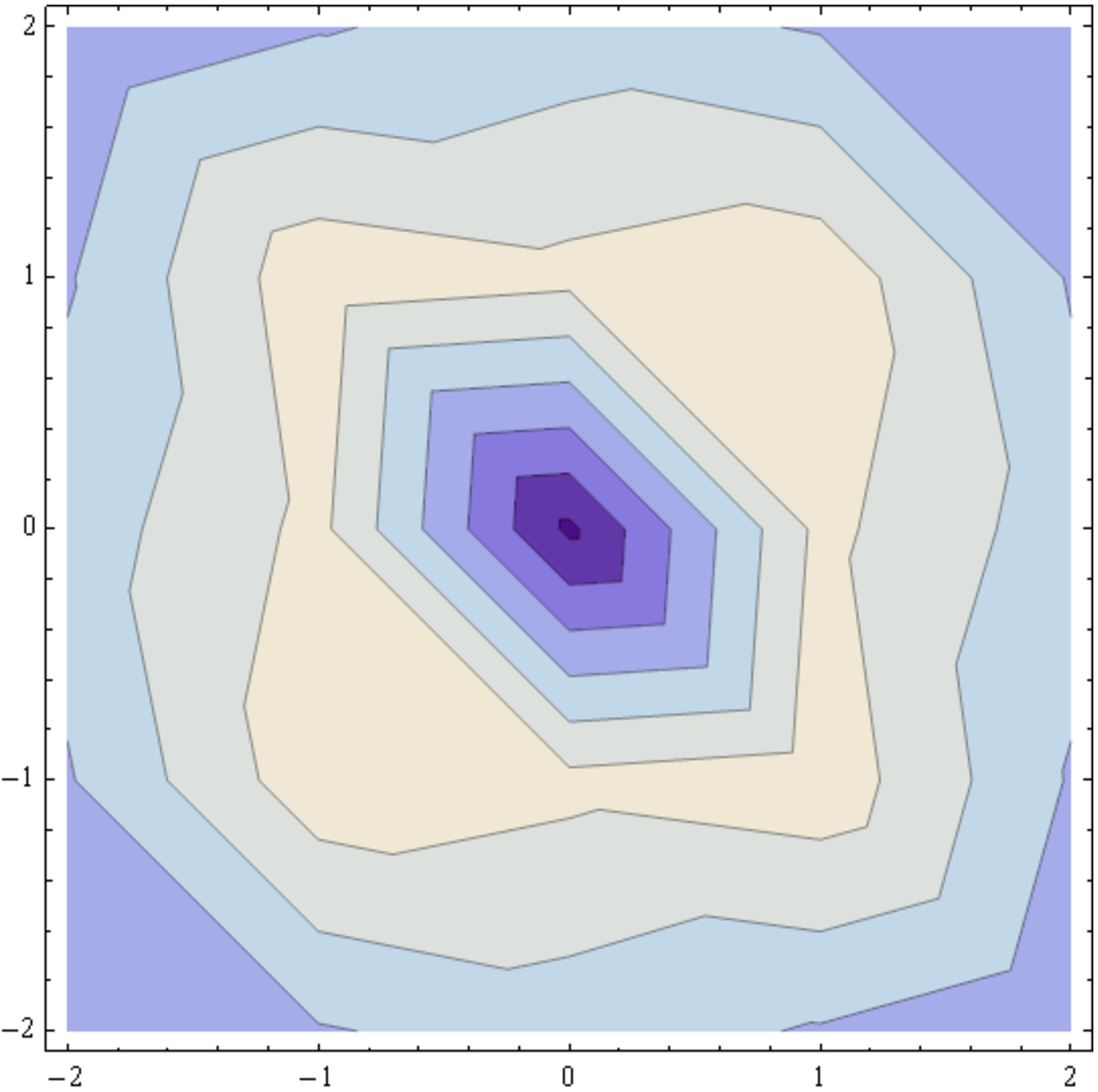} 
\hspace{.5cm}
\includegraphics[width=3.9cm]{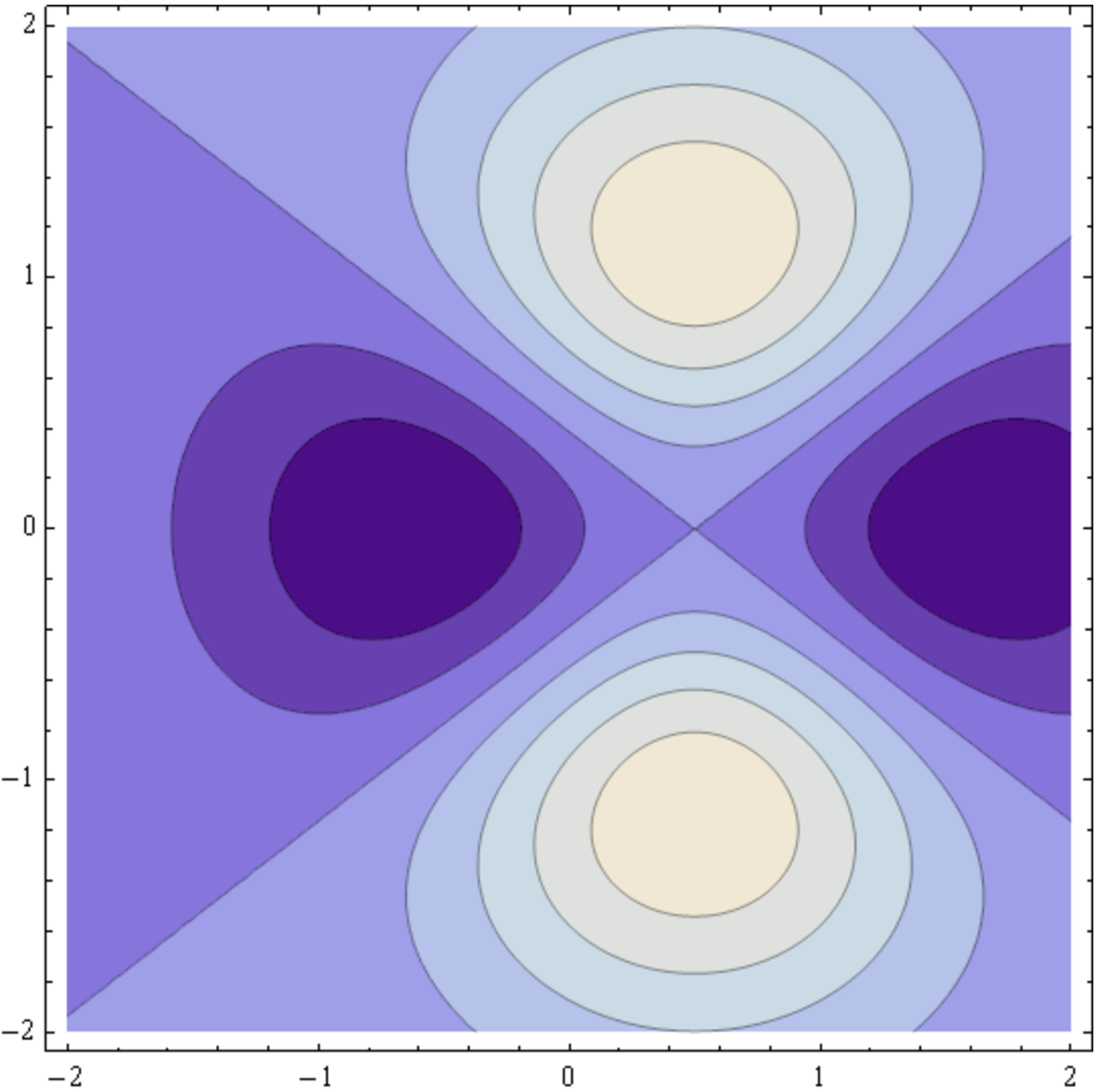}
\includegraphics[width=3.9cm]{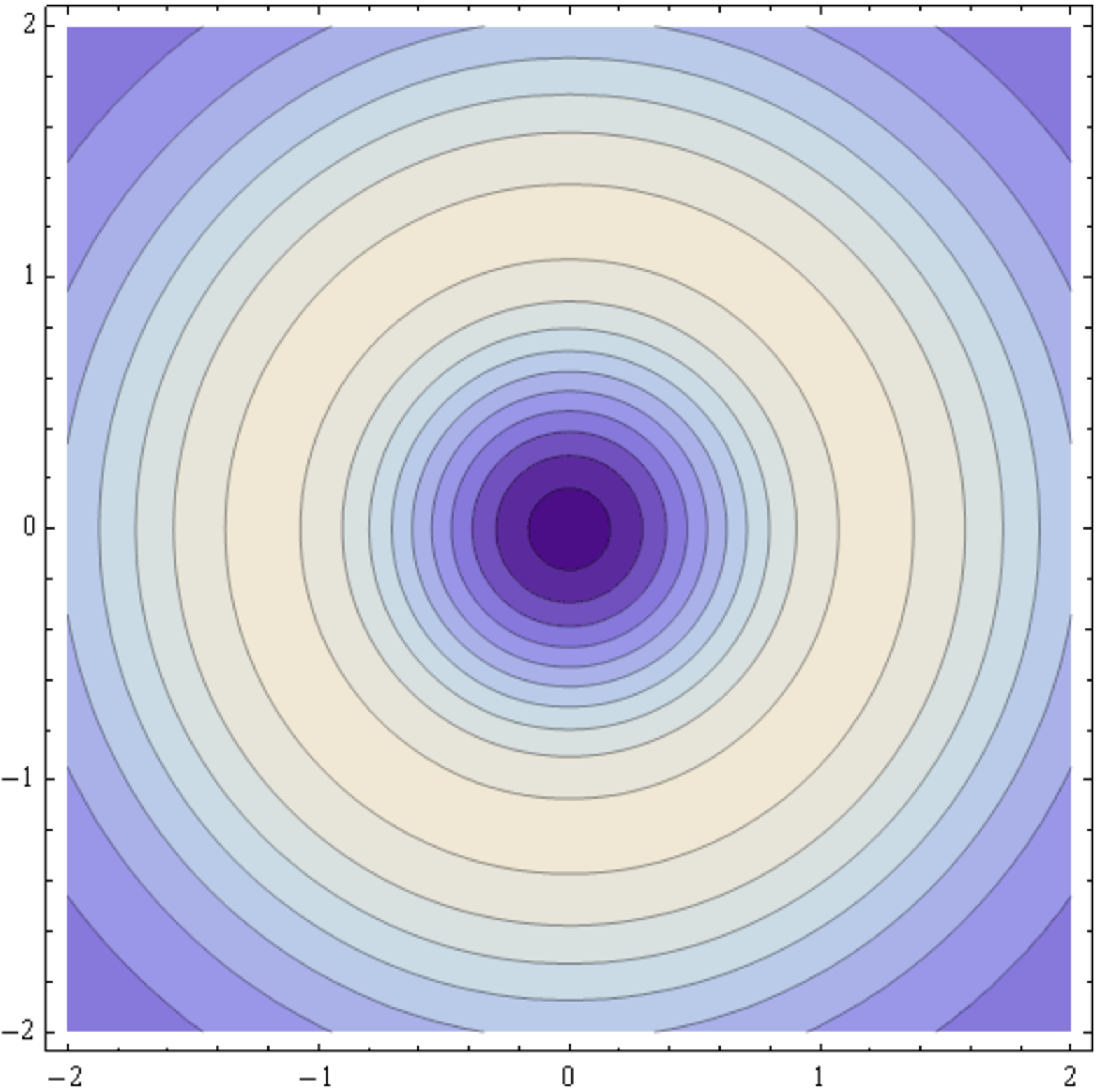}
\end{center}
\caption{Contour plots of $\delta K_{q,-q+p}$ for last low-lying mode in $p=(1,0,0,0)$ 
sector in four-dimensional case.
The left- and right-hand pairs of figures show \eq{eq:tensordk} from the numerical computations 
and \eq{eq:metricdk} substituted with \eq{eq:nonzeromixed}, respectively.
The left- and right-hand figures in each pair show $\delta K_{q,-q+p}$ at the slices $q=(q^1,q^2,0,0)$ and  
$q=(0,q^2,q^3,0)$, respectively.
}
\label{fig:dim4L2}
\end{figure}
In the four-dimensional case, I will show the numerical results for $L=2$ and $\alpha=1/L^2$. 
This number of $L$ is the maximum value feasible within the machine memory size\footnote{Even for
$L=2$, the precision of an array variable had to be made single to 
reduce the necessary memory size.}.
The low-lying spectra are plotted in Fig.~\ref{fig:dim4L2spec}.  
$|p|^4$ seems to be favored over $|p|^2$ as a fitting line. 

In more detail, in the $|p|=0$ sector, there are three types of spectrum,
$1.1\times 10^{-6}$, $1.8\times 10^{-4}$ and $6.5\times 10^{-4}$ with degeneracies of 
six, one and three, respectively. Therefore, these can consistently be identified with the modes 
\eq{eq:zerooff}, \eq{eq:zerodiag} and \eq{eq:zerotraceless}, respectively. 
In fact, the fitting of \eq{eq:metricdk} to \eq{eq:tensordk} from the numerical computations
was clearly consistent with the identification.
For example, an analysis of the second mode resulted in
$\delta g_{\mu\mu}=0.683 \pm 0.003$ and $\delta g_{\mu\nu}=0\pm 0.003$, which is
in clear agreement with \eq{eq:zerodiag}.

In each $|p|\neq 0$ sector, there are six low-lying spectra including their degeneracies,
which agrees with the result in the previous section.
For example, in the $p=(1,0,0,0)$ sector, there are three types of spectrum, 
$2.4\times 10^{-4}$, $3.5\times 10^{-4}$ and $5.1\times 10^{-4}$ with
degeneracies of two, three and one, respectively. Therefore these can be identified with
the modes \eq{eq:nonzerotraceless}, \eq{eq:nonzerooff} and \eq{eq:nonzeromixed}, respectively.
In fact, similar fitting procedures for each mode were consistent with this identification.
For example, an analysis of the last mode resulted in $\delta g_{11} =-0.41\pm 0.01,\ 
\delta g_{22,33,44}=0.80\pm 0.01$ and $\delta g_{\mu\nu}=0\pm 0.01$,
which can be well identified with \eq{eq:nonzeromixed}.
In Fig.~\ref{fig:dim4L2}, \eq{eq:tensordk} of the last mode 
and \eq{eq:metricdk} substituted with \eq{eq:nonzeromixed}
are shown. 

In summary, the numbers and properties of the low-lying modes of tensor models on fuzzy flat tori 
are numerically shown to be fully consistent with general relativity,
and the momentum dependence of the spectra is found to be $|p|^4$.
The latter fact suggests that the lowest-order effective action is composed of curvature square terms.

\section{Conclusion and future problems}
\label{sec:conclusion}

In this paper, I have considered a tensor model that possesses
Gaussian classical solutions corresponding to fuzzy flat spaces of arbitrary dimensions. 
The numerical analysis of the small fluctuations for two-, three- and four-dimensional fuzzy flat tori 
have shown clear agreement with general relativity on these tori. 
Therefore, it can be concluded 
that general relativity appears as an emergent phenomenon in this model.

The tensor models have a rank-three tensor as their only dynamical variable, 
and they are formulated background independently. 
Various background spaces can be generated as classical solutions
\cite{Sasakura:2005js,Sasakura:2005gv,Sasakura:2006pq}, and the 
above conclusion shows that small fluctuations around them can be identified with 
general relativity at least in a certain class of tensor models.
Thus, the tensor models can provide an interesting unifying description of space and geometry.
The specific choices of actions studied thus far must be generalized in future studies.

In the studies thus far, the actions were assumed to have certain semipositive definite forms 
and the classical solutions were assumed to be their absolute minima.
However, these assumptions force the effective actions to start with curvature square terms as is
consistent with the numerical analysis presented in this paper, since the Euclidean 
Einstein-Hilbert action is topological in two dimensions and generally takes both positive and 
negative values in dimensions higher than two \cite{Hamber:2007fk}. 
Therefore, indefinite actions or stationary solutions will be more 
interesting for the realization of the Einstein-Hilbert action, and will 
also provide further consistency checks in view of 
the qualitative difference between the two- and higher-dimensional cases.     

The next interesting question would be which fields can emerge in tensor models. 
Probably, the gauge fields can emerge from an emergent metric field
by a type of fuzzy analogue of the Kaluza-Klein mechanism. Then
how about fermions? In fact, tensor models have an interesting potential.
In the analyses performed thus far, the rank-three tensor is restricted to be totally symmetric,
which requires functions in fuzzy spaces to be commutative. By simply 
relaxing this restriction to another milder condition such as the generalized Hermiticity 
condition stated in refs.~\cite{Ambjorn:1990ge,Sasakura:1990fs,Godfrey:1990dt},
the functions can also become anticommutative, i.e., fermionic. 
These ideas suggest some interesting directions of future study in this area.

\section*{Acknowledgments}
The author was supported in part by a Grant-in-Aid for Scientific Research  
Nos.~16540244(C) and 18340061(B) from the Japan Society for the Promotion of Science (JSPS).


\end{document}